\documentclass[12pt]{article}

\title{\textbf{Algorithm for the maximum likelihood estimation of the parameters of the truncated normal and lognormal distributions}}

\author{Salvador Pueyo\thanks{E-mail: spueyo@ic3.cat}}

\date{}

\begin{document}

\maketitle

\begin{abstract}
\noindent This paper describes a simple procedure to estimate the parameters of the univariate truncated normal and lognormal distributions by maximum likelihood. It starts from a reparameterization of the lognormal that was previously introduced by the author and is especially useful when the lognormal is close to a power law, which is a limiting case of the first distribution. One of the new parameters quantifies the distance from the power law, and vanishes when the power law gives a sufficient description of the data. At this point, the other parameter equals the exponent of the power law. In contrast, when using the standard parameterization, the parameters of the lognormal diverge in the neighborhood of the power law. Whether or not we are in this neighborhood, the new parameters have properties that ease the process of estimation.

\end{abstract}

\section{Introduction}

Several different methods have been suggested for the estimation of the parameters of the truncated normal and lognormal distribution \cite{Koutroumbas2014}. This paper presents a simple algorithm for the estimation of these parameters in the univariate case. It starts from a reparameterization previously introduced by the author in some practical applications of the truncated lognormal in a complex systems context \cite{Pueyo2006b,Pueyo2013a}.

The lognormal distribution (whether truncated or not) can be expressed as \cite{Pueyo2006b}:
\begin{equation}
\label{Taylor}
\ln(f(x)) = \ln(a)-\beta\ln(x)-\psi[\ln(x)]^2,
\end{equation}
where $a$ is a constant that depends on $\beta$, $\psi$ and the truncation points if any. The equivalence to the standard parameters $\mu$ and $\sigma$ is \cite{Pueyo2006b}
\begin{equation}
\label{equivalencia}
\left\{ \begin{array}{lcl}
	\mu = -(\beta-1)/2\psi \\
	\sigma=1/\sqrt{2\psi}
\end{array} \right.
\end{equation}
Equation \ref{Taylor} is reduced to the equation of a power law distribution of exponent $\beta$ in the special case of $\psi=0$. This approximation is especially useful when the lognormal is in the neighborhood of a power law, i.e.\ when $\psi$ takes values close to zero, in which case $\mu$ and $\sigma$ diverge while $\beta$ and $\psi$ remain well-behaved. This can be the case when, as is likely in some empirical data  \cite{Pueyo2006b}, the lognormal results from small perturbations of a power law, rather than, e.g., from the central limit theorem. Then, $\psi$ quantifies the separation from the power law resulting from these perturbations  \cite{Pueyo2006b}. It is also the case when the lognormal distribution is used as an ingredient in the test of the goodness of fit to the power law in the method in \cite{Pueyo2013a}. In both contexts, the lognormal often appears truncated. A lower truncation can be unavoidable in empirical data because of limited resolution. An upper truncation can also occur in lognormals resulting from the perturbation of a power law, because, in empirical data, the later are often truncated \cite{Pueyo2011a}.

If we take $y=\ln(x)$, we obtain a normal distribution, which can be expressed, similarly to Eq.\ \ref{Taylor}, as
\[
\ln(f(y)) = a-\alpha y-\psi y^2,
\]
where
\begin{equation}
\label{alfabeta}
\alpha=\beta-1.
\end{equation}
Note that, in the limiting case of $\psi=0$, the normal becomes a negative exponential distribution of exponent $\alpha$.

Whether or not we are in the neighborhood of a power law (or of an exponential), these reparameterizations have interest because they suggest a simple algorithm to perform the maximum likelihood estimation (m.l.e.) of the parameters of the lognormal, or equivalently, of the normal, which is useful when these are truncated (whether from below, from above, or both), because, in this case, the standard parameters $\mu$ and $\sigma$ lose their straightforward relation to the expectation and standard deviation of the distribution.

I describe this method in the case of a set of data $\{y\}$ displaying a truncated normal distribution between $y_{min}$ and $y_{max}$:
\begin{equation}
\label{PDFy}
f(y) = \frac{e^{-\alpha y-\psi y^{2}}}{\displaystyle\int\limits_{y_{min}}^{y_{max}}e^{-\alpha u - \psi u^{2}}du}.
\end{equation}
These data $\{y\}$ might have been obtained by transforming logarithmically some given data $\{x\}$ that display a truncated lognormal spanning from $x_{min}=\exp(y_{min})$ to $x_{max}=\exp(y_{max})$.

\section{The algorithm}

By definition, the m.l.e.\ is performed by maximizing
\[
\Lambda = \displaystyle\sum\limits_{i=1}^{N} \ln(f(y_{i} | \alpha,\psi). 
\]
As usual, this can be done by seeking the values $\alpha$, $\psi$ that satisfy $\frac{\partial \Lambda}{\partial \alpha} = 0$, $\frac{\partial \Lambda}{\partial \psi} = 0$. In this case, this gives
\begin{equation}
\label{criteri}
\left\{ \begin{array}{lcl}
         E(y|\alpha,\psi)&=&\overline{y} \\
         E(y^{2}|\alpha,\psi)&=&\overline{y^{2}}
\end{array} \right.
\end{equation}
where $E$ is the expectation obtained from Eq.\ \ref{PDFy} and the overlines indicate sampling means. Since this does not have a closed form, the estimators have to be sought iteratively until finding the parameter values that satisfy Eq.\ \ref{criteri}.

There are several numerical methods to find extremes of arbitrary functions. However, I designed a special one for the truncated normal. 

Let us use the following notation for the update of the parameters at step $j+1$:
\begin{equation}
\label{recep1}
\left\{ \begin{array}{lcl}
\alpha_{j+1}&=&\alpha_{j}+\Delta_{j} \alpha\\
\psi_{j+1}&=&\psi_{j}+\Delta_{j} \psi
\end{array} \right.
\end{equation}
Equation \ref{criteri} suggests choosing a pair $(\Delta_{j} \alpha, \Delta_{j} \psi)$ that satisfies
\begin{equation}
\label{baseregla}
\left\{ \begin{array}{lcl}
\frac{\partial E(y)}{\partial \alpha} \Delta_{j} \alpha + \frac{\partial E(y)}{\partial \psi} \Delta_{j} \psi &=& \eta (\overline{y}-E(y| \alpha_{j}, \psi_{j}))\\
\frac{\partial E(y^{2})}{\partial \alpha} \Delta_{j} \alpha + \frac{\partial E(y^{2})}{\partial \psi} \Delta_{j} \psi &=& \eta (\overline{y^2}-E(y^{2}| \alpha_{j}, \psi_{j}))
\end{array} \right.
\end{equation}
for some $\eta : 0< \eta < 1$.

It follows from Eq.\ \ref{PDFy} that, in the neighborhood of the correct $\alpha$ and $\psi$:
\[
\left\{ \begin{array}{lcl}
\frac{\partial E(y)}{\partial \alpha}&=&-E(y^{2})+E(y)^{2}\\
\frac{\partial E(y)}{\partial \psi}&=&-E(y^{3})+E(y)E(y^{2})\\
\frac{\partial E(y^{2})}{\partial \alpha}&=&-E(y^{3})+E(y)E(y^{2})\\
\frac{\partial E(y^{2})}{\partial \psi}&=&-E(y^{4})+E(y^{2})^{2}
\end{array} \right.
\]
which can be approximated using the sampling moments:
\begin{equation}
\label{macarro}
\left\{ \begin{array}{lcl}
\frac{\partial E(y)}{\partial \alpha} & \approx & - \overline{y^{2}}+\overline{y}^{2}\\
\frac{\partial E(y)}{\partial \psi} & \approx & -\overline{y^{3}} + \overline{y} \overline{y^{2}}\\
\frac{\partial E(y^{2})}{\partial \alpha} & \approx & -\overline{y^{3}}+\overline{y}\overline{y^{2}}\\
\frac{\partial E(y^{2})}{\partial \psi} & \approx & -\overline{y^{4}}+\overline{y^{2}}^{2}\\
\end{array} \right.
\end{equation}

The updating rule that results from Eqs.\ \ref{baseregla}-\ref{macarro} is:
\begin{equation}
\label{recep2}
\left\{ \begin{array}{lcl}
\Delta_{j} \alpha &=& a\eta (\overline{y}-E(y| \alpha_{j}, \psi_{j}))+b\eta (\overline{y^2}-E(y^{2}| \alpha_{j}, \psi_{j}))\\
\Delta_{j} \psi &=& b\eta (\overline{y}-E(y| \alpha_{j}, \psi_{j}))+c\eta (\overline{y^2}-E(y^{2}| \alpha_{j}, \psi_{j}))
\end{array} \right.
\end{equation}
where
\begin{equation}
\label{recep3}
\left\{ \begin{array}{lcl}
a &=& (\overline{y^{4}}-\overline{y^{2}}^{2})/h\\
b &=& (-\overline{y^{3}} + \overline{y} \overline{y^{2}})/h\\
c &=& (\overline{y^{2}}-\overline{y}^{2})/h
\end{array} \right.
\end{equation}
and
\begin{equation}
\label{recep4}
h=\overline{y^{4}}(-\overline{y^{2}}+\overline{y}^{2})+\overline{y^{3}}(\overline{y^{3}}-2\overline{y}\overline{y^2})+\overline{y^2}^{3}.
\end{equation}

The algorithm thus consists of Eq.\ \ref{recep1} and Eqs.\ \ref{recep2}-\ref{recep4}. The expectations in Eq.\ \ref{recep2} are calculated from Eq.\ \ref{PDFy}, using numerical integraton.

The value of $\eta$ is relatively arbitrary. The larger the chosen $\eta$, the quicker the convergence if it converges, but also the larger the risk that it does not. The author generally uses $\eta = 0.33$. In case of overflow, $\eta$ should be reduced. As usual, the algorithm should proceed until $\Delta_{j} \alpha$ and $\Delta_{j} \psi$ become smaller than some given thresholds, which determine the precission of the estimates.

Once we have estimated $\alpha$ and $\psi$, it is straightforward to obtain $\mu$ and $\sigma$ from Eqs.\ \ref{equivalencia}-\ref{alfabeta}.

\bibliographystyle{unsrt}

\end{document}